\documentclass[%
reprint,
showkeys,
showpacs,preprintnumbers,
amsmath,amssymb,
aps,
pra,
floatfix,
onecolumn]{revtex4-2}
\usepackage{appendix}
\usepackage{amsmath,latexsym}
\usepackage{scalerel}
\usepackage{hyperref}
\usepackage{mathtools}
\usepackage{newtxtext,newtxmath}
\usepackage{setspace}
 
\begin{document}
	\title{Emergence of cosmic space in Tsallis modified gravity from equilibrium and non-equilibrium thermodynamic perspective}
	\author{M. Dheepika}
	\email{mdheepika@cusat.ac.in}
	\author{Hassan Basari V. T.}
	\email{basari@cusat.ac.in}
	\author{Titus K. Mathew}
	\email{titus@cusat.ac.in}
	\affiliation{Department of Physics, Cochin University of Science and Technology, Kochi, Kerala 682022, India}
	\begin{abstract}{\vspace{1.0cm}\textbf{Abstract.}}
		In this paper, we obtain the law of emergence with Tsallis entropy from the thermodynamic laws. We first derive the law of emergence from the equilibrium description of the unified first law and Clausius relation. However, it has been shown that considering Tsallis entropy as the horizon entropy, the Clausius relation $\delta Q=T dS$ 
		does not hold due to non-equilibrium thermodynamics and is replaced by the entropy-balance relation $dS=\frac{\delta Q}{T}+d_{i}S$, where $d_iS$ is the additional entropy produced due to the irreversible thermodynamic process. Hence, we derive the law of emergence from the non-equilibrium description of thermodynamic laws. The comparison between the law of emergence in both cases shows that the law of emergence from the non-equilibrium approach describes the expansion of the universe in terms of the areal volume, unlike the effective volume in the equilibrium case. We have further shown that the law of emergence also satisfies the condition of the maximization of entropy for a de Sitter universe; thus, the entropy of the universe evolves to a bounded value in the asymptotic future.
		
	\end{abstract}

	\maketitle
	
	\newpage
\section{Introduction}
The developments in black hole thermodynamics in the last decades of the twentieth century led to the discovery of the profound connection between horizon thermodynamics and gravity. Hawking~\cite{PhysRevLett.26.1344} took the first step and showed that when two black holes merged, the horizon area of the resulting black hole would be greater than the sum of the horizon areas of the individual black holes. Following this, Bekenstein~\cite{bekenstein1972black,PhysRevD.7.2333,PhysRevD.9.3292} put forth that the area of the black hole horizon can be taken as a measure of its entropy. Meanwhile, the four laws of black hole dynamics were formulated~\cite{Bardeen1973} and found to be analogous to the laws of thermodynamics. Later, Hawking~\cite{HAWKING1974,Hawking1975} discovered that black holes possess temperature, which is proportional to surface gravity. Soon after, Davies~\cite{Davies_1975} and Unruh~\cite{PhysRevD.14.870} generalized Hawking's idea to any observer accelerating in a flat space-time and showed that the horizon perceived by such an observer could have a temperature proportional to the observer's acceleration. In extending this idea to cosmology, Gibbons and Hawking~\cite{PhysRevD.15.2738} have shown that the horizon of an expanding universe possesses both entropy and temperature. Taking account of these results, Jacobson~\cite{PhysRevLett.75.1260} derived the Einstein field equation from the Clausius relation on a local Rindler horizon, assuming the Bekenstein-Hawking area law of entropy. These results motivate further research on the connection between the fundamental laws of thermodynamics and gravity.  

Padmanabhan~\cite{Padmanabhan_2002} restructured Einstein's equation of gravity as a thermodynamic identity on a spherically symmetric space-time. This result indicates that Einstein's equation describes the thermodynamics of space-time~\cite{Padmanabhan:2002xm,Padmanabhan2002Dec,Padmanabhan2003Jul,Padmanabhan2003Dec,PADMANABHAN200549}. Later, Paranjape et al.~\cite{PhysRevD.74.104015} extended this result to more general  Lanczos-Lovelock gravity. This result firmly pointed out that the thermodynamic interpretation of the field equation can be valid even with the inclusion of quantum corrections to the Einstein-Hilbert action since the higher derivative terms in Lanczos-Lovelock gravity can be treated as the quantum corrections to general relativity~\cite{KOTHAWALA2007338}. At around the same time, Cai and Cao~\cite{PhysRevD.75.064008} showed that the Clausius relation holds for Lovelock gravity, where they adopt the corresponding black hole entropy formula for the apparent horizon and treat the higher derivative terms in Lovelock theory as an effective energy-momentum tensor. 

The thermodynamic connection of Einstein's equation of gravity has implications in cosmology. Cai and Kim~\cite{Cai_2005} obtained the Friedmann equation, from the Clausius relation, in Einstein's gravity using the area law of entropy. These authors extended the results to Gauss-Bonnet and Lovelock gravity theories using the black hole entropy in respective gravity theories. Later Akbar and Cai~\cite{PhysRevD.75.084003} restructured the Friedmann equation into the first law of thermodynamics in Einstein, Gauss-Bonnet, and Lovelock gravity theories. However, the same authors~\cite{AKBAR20067} later found it impossible to obtain the Friedmann equations in scalar-tensor gravity and $f(R)$ theory from the Clausius relation using the corresponding black hole entropy relation. Nevertheless, they could obtain the Friedmann equations in these gravity theories from the Clausius relation by defining an effective energy-momentum tensor from the modified field equation and the conventional area law of entropy. 

A significant turn in treating the horizon thermodynamics in gravity theories with higher-order curvature contributions to entropy was brought by Eling et al.~\cite{PhysRevLett.96.121301}. The authors suggested that changing the perspective from equilibrium to non-equilibrium thermodynamics is essential while considering the higher-order curvature corrections to the entropy for the successful derivation of the field equation. Following this intriguing idea, Akbar and Cai~\cite{AKBAR2007243} have rewritten the $f(R)$ gravity field equation in the FRW setup into the non-equilibrium description of the first law of thermodynamics. Later Cai and Cao~\cite{PhysRevD.75.064008} showed that scalar-tensor gravity requires a non-equilibrium thermodynamic perspective. These results indicate that theories of gravity having entropy with higher order curvature contributions or other than area law form for black hole entropy might require a non-equilibrium thermodynamic perspective.

The deep connection between gravity and thermodynamics directly implies that the space-time continuum is thermodynamic and can carry heat like a fluid~\cite{https://doi.org/10.48550/arxiv.0910.0839,Padmanabhan2014}. It is to be noted that thermodynamics is an emergent phenomenon that deals with the connection of macroscopic variables like pressure and temperature, which 
has no existence in the microscopic domain. Following this, Padmanabhan proposed that gravity could be an emergent phenomenon. Accordingly, macroscopic geometrical properties like metric and curvature have no existence in the microscopic realm. In extending the emergent paradigm to cosmology, Padmanabhan, in 2012~\cite{padmanabhan2012emergence,Padmanabhan_2012_Aug}, proposed that ``cosmic space is emergent as cosmic time progresses''. According to this, the expansion of the universe is happening
due to the difference in the degrees of freedom on the horizon, $N_{\text{sur}}$, and that on the bulk enclosed by the horizon, $N_{\text{bulk}}$. As the universe expands, the difference in these degrees of freedom will decrease and finally attain zero, at which it achieves a static equilibrium.  This equality between the degrees of freedom
at the end stage is known as the holographic equipartition condition. Thus the emergence of space is the quest of the universe for achieving holographic equipartition condition. Accordingly, Padmanabhan~\cite{padmanabhan2012emergence} proposed a fundamental principle, known as the law of emergence, to describe the expansion of the universe as $\frac{dV}{dt}=L_p^2(N_{\text{sur}}-N_{\text{bulk}})$, where $V$ is the Hubble volume of a flat $(3+1)$-dimensional FRW universe and $t$ is the cosmic time. Padmanabhan~\cite{padmanabhan2012emergence} obtained the Friedmann equation from the above law of emergence in the spatially flat universe in the context of Einstein's gravity. 

Padmanabhan's idea was extended to $(n+1)$ Einstein's gravity, Gauss-Bonnet, and Lovelock gravity by proposing the law of emergence for a flat FRW universe, and the corresponding Friedmann equations were obtained~\cite{Cai2012}. The author used a modified expression for surface degrees of freedom to propose the law of emergence in $(n+1) $ Einstein's gravity for a flat universe. In extending the law to Gauss-Bonnet and Lovelock gravity theories, the same authors have used an effective volume, different from the areal volume, in addition to the modified expression for the surface degrees of freedom. On the other hand, Yang et al.~\cite{PhysRevD.86.104013} proposed an alternative generalization of the law of emergence, such that the increase in horizon volume is taken to be proportional to a function, $f(\Delta N, N_{\text{sur}})$, where $\Delta N=N_{\text{sur}}-N_{\text{bulk}}.$ Even though this generalization retains the horizon volume as the areal volume, the function $f(\Delta N, N_{\text{sur}})$ has a complicated form except in Einstein's gravity. Later Sheykhi~\cite{PhysRevD.87.061501} extended Padmanabhan's proposal to a universe of $(n+1)$ dimension with any spatial curvature in the context of Einstein and more general Gauss-Bonnet and Lovelock gravities and derived the corresponding Friedmann equations. Subsequently, Yuan and Huang~\cite{https://doi.org/10.48550/arxiv.1304.7949} obtained the modified Friedmann equation from the law of emergence with logarithmic and power-law corrected entropy relations. Later Eune and Kim~\cite{PhysRevD.88.067303} extended Padmanabhan's idea by considering the invariant volume of the horizon in a non-flat universe. However, this work has been criticized~\cite{Mahith_2018,Hareesh_2019} for using a time-dependent Planck length. Meantime, Ali~\cite{FARAGALI2014335} modified the law of emergence using a generic form of entropy and derived the Friedmann equation in $(3+1)$ Einstein's gravity and Gauss-Bonnet gravity. While Ai et al.~\cite{PhysRevD.88.084019} proposed another modified form for the law of emergence, in which they used the basic relation $N_{\text{sur}}=4S$ and thus expressed the rate of change of the horizon volume as proportional to $\frac{1}{H} \frac{dN_{\text{sur}}}{dt}$. Later, Young and Lee~\cite{Chang-Young2014} pointed out that, for the non-flat universe, the Friedmann equation emerges from Padmanabhan's proposal only when the areal volume is used and argued that the holographic principle gives a clue about the flatness of our universe. The literature also shows the application of the idea of the emergence of space to braneworld scenarios~\cite{Sheykhi_2013,SHEYKHI201323} and the derivation of the corresponding cosmological equations. 

Later on, Dezaki and Mirza~\cite{Dezaki2015} introduced the connection between the law of emergence and thermodynamics by obtaining the law of emergence for a $(3+1)$ non-flat FRW universe in Einstein's gravity from the first law of thermodynamics. However, a more direct method of obtaining the law of emergence in $(n+1)$ Einstein, Gauss-Bonnet, and Lovelock gravities, from the unified first law of thermodynamics has been proposed by Mahith et al.~\cite{Mahith_2018}. Another meaningful thermodynamic connection to the law of emergence was obtained later by Krishna and Mathew~\cite{PhysRevD.96.063513,PhysRevD.99.023535,https://doi.org/10.48550/arxiv.2002.02121}. The authors found that the emergence of cosmic space can be regarded as a tendency to maximize the horizon entropy by investigating the relationship between the emergence of space and the entropy maximization in the context of Einstein, Gauss-Bonnet, and Lovelock gravities. Meanwhile, Hareesh et al.~\cite{Hareesh_2019} found the reason for preferring areal volume in obtaining the law of emergence from either the Clausius law or the unified first law. It has been shown that the Clausius relation and the unified first law are defined properly in a non-flat universe, only with areal volume. Recently, Basari et al.~\cite{https://doi.org/10.48550/arxiv.2111.00726,https://doi.org/10.48550/arxiv.2209.00304} introduced a generic derivation of the law of emergence in gravity theories with entropy different from the area law of entropy like Gauss-Bonnet and Lovelock gravity from the thermodynamic laws. The same idea is then extended to $f(R)$ gravity~\cite{https://doi.org/10.48550/arxiv.2111.00726}, where the authors used a non-equilibrium thermodynamic perspective as the theory demands it to obtain the law of emergence. 

Until now, we summarized the various aspects of the connection between gravity and thermodynamics and the different approaches describing the emergence of space that explains the expansion of the universe. Now we will motivate the use of non-equilibrium thermodynamics in the approach to the emergence of space in cases where one has to consider other than the Bekenstein-Hawking entropy. 
In a given gravity theory, one needs to consider the corresponding entropy of the black hole to obtain the law of emergence of cosmic space from the fundamental thermodynamic laws. In modified gravity theories, the entropy of black holes might differ from the area law of entropy. It is also possible to assume some general entropy form and analyze its cosmological implication. A need for a generalized entropy for the complex system came in 1902 when Gibbs pointed out that the Boltzmann-Gibbs theory cannot be used for complex systems like gravitational systems whose partition function diverges~\cite{Tsallis2013}. Later in 1988, Tsallis~\cite{Tsallis1988} proposed a generalization to the Boltzmann-Gibbs entropy for the multifractal structures. Following this, in 2013, Tsallis and Cirto~\cite{Tsallis2013} proposed that the entropy of the complex systems, whose elements are strongly correlated through long-range interactions or strong quantum entanglement, like black holes should be identified with the generalized non-extensive entropy, referred to as Tsallis entropy. 

Inquisitive to know the implications of the non-extensive entropy in cosmology, Sheykhi~\cite{SHEYKHI2018118} assumed Tsallis entropy, $S=\gamma A^{\beta},$ where $\gamma$ is an unknown constant and $\beta$ is a constant termed as the non-extensive parameter for the horizon and derived the Friedmann equations from the unified first law of thermodynamics, for an FRW universe with any spatial curvature. In this derivation, the author considers the cosmic components as perfect fluids that obey the standard conservation law. Further, the author proposed a modified law of emergence and obtained the same Friedmann equations from it. As an extension, Chen~\cite{Chen2022} proposed the law of emergence with Tsallis entropy for the $(n+1)$-dimensional non-flat FRW universe and obtained the corresponding Friedmann equations. They also obtained the same Friedmann equations from the unified first law by using Tsallis entropy. Recently through a unified expansion law derived from the equilibrium first law of thermodynamics, Basari et al.~\cite{https://doi.org/10.48550/arxiv.2209.00304} obtained the law of emergence for the Tsallis entropy given by $S=\frac{A_{0}}{4L^{n-1}_{\scaleto{P}{3.5pt}}}\left(\frac{A}{A_{0}}\right)^{\beta}$, where $A_{0}$ is a constant and $L^{n-1}_{\scaleto{P}{3.5pt}}$ is the Planck length. Earlier to this,  Lymperis and Saridakis~\cite{Lymperis2018} extracted the modified Friedmann equations, using Tsallis entropy of the form $S=\frac{\tilde{\alpha}}{4G}A^{\beta}$ where $\tilde{\alpha}$ is an unknown constant, from the equilibrium Clausius relation. They have extracted an effective dark energy and rewritten the modified Friedmann equation in the standard form, similar to that in (3+1) Einstein's gravity, and studied cosmology. Nojiri et al.~\cite{Nojiri2019} also extracted the modified Friedmann equations from the equilibrium Clausius relation using Tsallis entropy given by $S=\frac{A_{0}}{4G}\left(\frac{A}{A_{0}}\right)^{\beta}$, where $A_{0}$ is an unknown constant but the exponent $\beta$ as the varying, and studied the evolution in late and early phase. Later in reference~\cite{NOJIRI2020114850}, Nojiri et al. restructured the modified Friedmann equation, which is obtained using the previous form of Tsallis entropy into a form similar to the standard Friedmann equation by defining an effective fluid, which is equivalent to dark energy and explained the late acceleration.

In the above works, the authors have obtained the Friedmann equations using Tsallis entropy, either directly from the law of thermodynamics or by proposing the law of emergence, assuming the conventional conservation law. However, the validity of the equilibrium first law of thermodynamics, say the Clausius relation with the standard energy-momentum conservation, is doubtful in the case of entropies other than Bekenstein-Hawking entropy. Eling et al.~\cite{PhysRevLett.96.121301} have revealed the contradiction between the equilibrium Clausius relation and the standard energy-momentum conservation in the case of entropy with higher-order curvature corrections. 
The authors resolved it by adding an entropy production term $d_{i}S$, which is the additional entropy generated due to the irreversible processes, to the entropy balance relation. Resulting non-equilibrium entropy balance relation, $dS = \frac{\delta Q}{T} + d_{i}S$ is used to obtain the field equation by considering a divergence-free matter stress tensor. 
Following this approach, Asghari and Sheykhi~\cite{10.1093/mnras/stab2671} have shown that, with Tsallis entropy, a non-extensive entropy, one needs the non-equilibrium thermodynamic description to restore the standard energy-momentum conservation. These authors used the non-equilibrium entropy balance relation with Tsallis entropy and derived the field equation by considering a divergence-free matter stress tensor. Further, they have obtained the corresponding  Friedmann equations, which differ from the one obtained by Sheykhi~\cite{SHEYKHI2018118} with equilibrium thermodynamics using Tsallis entropy.

As mentioned earlier, the law of emergence with Tsallis entropy has been proposed, and the Friedmann equations are derived in the context of equilibrium thermodynamics. However, Tsallis entropy demands a non-equilibrium treatment. Hence obtaining the law of emergence in the context of non-equilibrium thermodynamic description is necessary. That motivates us to obtain the law of emergence using the Tsallis entropy from the non-equilibrium thermodynamic perspective. We accomplish our purpose using a unified formulation of Tian and Booth~\cite{PhysRevD.90.104042}, which was proposed to derive the Friedmann equations from non-equilibrium thermodynamics in the context of modified gravity theories. Unlike the previous works in Tsallis entropy, we first restructure the Tsallis entropy $S=\gamma A^{\beta}$ into a form, $S=\frac{A}{4G_{\scaleto{\text{eff}}{3.5pt}}}$, similar to Bekenstein-Hawking entropy, where $G_{\scaleto{\text{eff}}{3.5pt}}$ firmly depends on the nature of the parameter $\gamma$. The $G_{\scaleto{\text{eff}}{3.5pt}}$ used here is different from the idea of varying gravitational constant in Dirac's large numbers hypothesis~\cite{10.2307/78591}. The field equation corresponding to the above entropy form can  be expressed in a form, $G_{\mu \nu}=8\pi G_{\scaleto{\text{eff}}{3.5pt}}T^{\text{(eff)}}_{\mu \nu}$ which is similar to the standard field equation in Einstein's gravity. The covariant derivative of this field equation results in a continuity equation $\dot{\rho}_{\scaleto{\text{eff}}{3.5pt}}+nH(\rho_{\scaleto{\text{eff}}{3.5pt}}+P_{\scaleto{\text{eff}}{3.5pt}})=-\frac{\dot{G_{\scaleto{\text{eff}}{3.5pt}}}\rho_{\scaleto{\text{eff}}{3.5pt}}}{G_{\scaleto{\text{eff}}{3.5pt}}}$. Unlike the conventional continuity equation for a perfect fluid, a non-zero term appears on the right-hand side of the above equation, which effectively balances the energy flow due to the non-equilibrium situation. The corresponding
non-equilibrium energy dissipation, $\mathcal{E},$ causes the generation of an additional entropy, $d_{\scaleto{P}{3.5pt}}S.$ Taking account of these additional energy and entropy, the Clausius relation and the unified first law of thermodynamics modifies to $\delta Q=\tilde{T}_{\scaleto{A}{3.5pt}}(dS+d_{\scaleto{P}{3.5pt}}S)$ and $dE=A\psi+WdV+\mathcal{E}$ respectively, which then constitute the non-equilibrium thermodynamic relations. From these non-equilibrium thermodynamic relations, we obtain the law of emergence. We also check whether the resulting law of emergence implies entropy maximization. 

This paper is organized as follows. In the forthcoming session, we derive the law of emergence from the laws of thermodynamics through the equilibrium approach. In section 3, we obtain the law of emergence from the unified first law of thermodynamics and Clausius relation in the non-equilibrium perspective and also see whether the law indicates entropy maximization. In section 4, we conclude our results.

\section{Emergence of cosmic space in Tsallis modified gravity from equilibrium perspective}
In this section, we obtain the law of emergence with Tsallis entropy in the context of equilibrium thermodynamics.
The spatially homogeneous and isotropic $(n+1)$-dimensional universe is represented by the line element~\cite{Bak_2000}
\begin{equation}\label{n+1m}
	ds^{2}=h_{\mu \nu}dx^{\mu}dx^{\nu}+\tilde{r}^{2}d\Omega^{2}_{n-1},
\end{equation}
where $h_{\mu \nu}=diag(-1,\frac{a^2(t)}{1-kr^2})$ is the 2-dimensional metric with coordinates $x^{0}=t$ and $x^{1}=r$, $\tilde{r}=a(t)r$, $a(t)$ is the scale factor, $k$ is the curvature index, and $d\Omega^{2}_{n-1}$ denotes the line element of $(n-1)$-dimensional unit sphere.  We consider the universe as a thermodynamic system bounded by the apparent horizon, which satisfies the relation, $h^{\mu \nu}\partial_{\mu}\tilde{r}\partial_{\nu}\tilde{r}=0$, and hence the radius is given by~\cite{Bak_2000}
\begin{equation}\label{ah}
	\tilde{r}_{\scaleto{A}{3.0pt}}={\frac{1}{\sqrt{H^2+\frac{k}{a^2}}}},
\end{equation}
where $H$ is the Hubble parameter. The temperature of the apparent horizon~\cite{https://doi.org/10.48550/arxiv.1502.04235,https://doi.org/10.48550/arxiv.1505.07371} is proportional to surface gravity, $\kappa$, and is given by the expression 
\begin{equation}\label{aht}
	T_{\scaleto{A}{3.5pt}}=-\frac{\kappa}{2 \pi}=\frac{1}{2\pi \tilde{r}_{\scaleto{A}{3.0pt}}} \left( 1-\frac{ \dot{\tilde{r}}_{\scaleto{A}{3.5pt}}}{2H\tilde{r}_{\scaleto{A}{3.0pt}}} \right) ,
\end{equation}
where over-dot represents the derivative with respect to cosmic time. We use units with $\hbar=1=c=k$ and Planck length $L_p^2=G$, where $G$ is the gravitational constant. Here we consider the generalized non-extensive Tsallis entropy, which is the entropy of large gravitating systems, for the apparent horizon, which is given by the expression~\cite{TAVAYEF2018195}
\begin{equation}\label{ahs}
	S_{\scaleto{A}{3.5pt}}=\gamma A^{\beta},
\end{equation}
where $\gamma$ is a positive unknown constant~\cite{Moradpour2016}, $\beta$ is the Tsallis parameter, $A=n\Omega_{n}\tilde{r}^{n-1}_{\scaleto{A}{3.5pt}}$ is the area of the apparent horizon, and $\Omega_{n}=\frac{\pi^{\frac{n}{2}}}{\Gamma(\frac{n}{2}+1)}$ is the volume of an $n$-dimensional unit sphere~\cite{Cai_2005}. When $\beta=1$ and $\gamma=\frac{1}{4G}$, 
Tsallis entropy reduces to Bekenstein-Hawking entropy. In literature~\cite{SHEYKHI2018118,Lymperis2018,Nojiri2019,NOJIRI2020114850,10.1093/mnras/stab2671}, different authors have assumed slightly different forms for the proportionality constant $\gamma$ for convenience.  However, in general, $\gamma$ is assumed to be inversely proportional to the gravitational constant $G$. Since we are considering the equilibrium perspective, the matter and energy content of the universe are considered to be in the form of perfect fluid of energy density $\rho$ and pressure $P$, satisfying the continuity equation
\begin{equation}\label{ce}
	\dot{\rho}+nH(\rho+P)=0.
\end{equation}
Following these, we derive the law of emergence from the equilibrium unified first law of thermodynamics and Clausius relation in the subsequent subsections.
\subsection{Emergence of cosmic space from the unified first law of thermodynamics} 
\label{sec:1A}
The universal thermodynamic identity, the unified first law of thermodynamics~\cite{Hayward_1998}, is given by the expression
\begin{equation}\label{UFL}
	dE=A\psi+WdV,
\end{equation}
where $W=(1/2)(\rho - P)$ is the work density, $E=\rho V$ is the total energy content inside the horizon of radius $\tilde{r}_{\scaleto{A}{3.0pt}}$,  and volume, $V=\Omega_{n}\tilde{r}^{n}_{\scaleto{A}{3.5pt}}.$ Here $\psi$ is the energy flux density,  which is defined as the vector $\psi_{\mu}=T_{\mu}^{\nu}\partial_{\nu}\tilde{r}+W\partial_{\mu} \tilde{r}$~\cite{Hayward_1998,PhysRevD.75.064008,PhysRevD.90.104042}, where $T_{\mu}^{\nu}$ is the energy-momentum tensor projected onto the two-dimensional space-time normal to the $(n-1)-$dimensional sphere. For a perfect fluid, it is given by~\cite{PhysRevD.75.064008,PhysRevD.90.104042}
\begin{align}\label{EFD}
	\psi&=\psi_{t}dt+\psi_{\tilde{r}_{\scaleto{A}{2.5pt}}}d\tilde{r}_{\scaleto{A}{3.0pt}}  
	=-(\rho+P)H\tilde{r}_{\scaleto{A}{3.0pt}}dt+\frac{1}{2}(\rho+P)d\tilde{r}_{\scaleto{A}{3.0pt}}.
\end{align}
This determines the total energy flow through the apparent horizon. Here $\psi_{t}$ and $\psi_{\tilde{r}_{\scaleto{A}{2.5pt}}}$ are the time part and space part of  $\psi$. Following the Clausius relation, the entropy $S_{\scaleto{A}{3.5pt}},$ associated with the heat flow $\delta Q$, through the apparent horizon during the infinitesimal time interval, $dt$, is
\begin{equation}\label{CR1}
	\delta Q=\tilde{T}_{\scaleto{A}{3.5pt}}dS_{\scaleto{A}{3.5pt}}.
\end{equation} 
During this infinitesimal interval of time, the horizon is assumed to be stationary; hence, its radius is fixed. Therefore the temperature of the horizon during this interval is taken to be  $\tilde{T}_{\scaleto{A}{3.5pt}}=\frac{1}{2\pi \tilde{r}_{\scaleto{A}{3.0pt}}}.$  Here $\delta Q$ is equivalent to $\left.-dE\right|_{\tilde{r}_{\scaleto{A}{2.5pt}}=\text{constant}}=-A\psi_{t}dt,$ the amount of energy crossing the apparent horizon during the infinitesimal time interval. Hence we have,
\begin{equation}\label{CR2}
	\tilde{T}_{\scaleto{A}{3.5pt}}dS_{\scaleto{A}{3.5pt}}=\frac{1}{2\pi \tilde{r}_{\scaleto{A}{3.0pt}}}dS_{\scaleto{A}{3.5pt}}=-A\psi_{t}dt.
\end{equation}
Substituting equations~\eqref{EFD},~\eqref{CR2} and the work density $W$ in~\eqref{UFL}, we can express the unified first law~\cite{PhysRevD.92.024001} as
\begin{align}\label{UFLTS1}
	dE&=-\frac{1}{2\pi \tilde{r}_{\scaleto{A}{3.0pt}}} dS_{\scaleto{A}{3.5pt}}+\rho dV.
\end{align}
Let us now multiply the above equation~\eqref{UFLTS1} throughout by  $\left( 1-\frac{\dot{\tilde{r}}_{\scaleto{A}{3.5pt}}}{2H\tilde{r}_{\scaleto{A}{3.0pt}}} \right) $, and on simplification will leads to,
\begin{equation}\label{UFLTS2}
	dE=-\frac{1}{2\pi \tilde{r}_{\scaleto{A}{3.0pt}}}\left( 1-\frac{ \dot{\tilde{r}}_{\scaleto{A}{3.5pt}}}{2H\tilde{r}_{\scaleto{A}{3.0pt}}} \right)dS_{\scaleto{A}{3.5pt}}+WdV=-T_{\scaleto{A}{3.5pt}}dS_{\scaleto{A}{3.5pt}}+WdV.
\end{equation}
For pure de Sitter universe, $\rho=-P,$ for which the above equation reduces to  $dE=-T_{\scaleto{A}{3.5pt}}dS_{\scaleto{A}{3.5pt}}-PdV.$ 

We will now proceed to derive the law of emergence with Tsallis entropy from the unified first law of thermodynamics given in equation~\eqref{UFLTS2}. Substituting the expression for Tsallis entropy~\eqref{ahs} for the horizon entropy and using continuity equation~\eqref{ce}, the first law in~\eqref{UFLTS2} reduces to
\begin{equation}\label{UFL_LOE:EQ2}
	\tilde{r}^{(n-1)\beta-n-2}_{\scaleto{A}{3.5pt}}d \tilde{r}_{\scaleto{A}{3.0pt}}=-\frac{2\pi \Omega_{n}d\rho}{\gamma \beta (n \Omega_{n})^{\beta}(n-1)}.
\end{equation}
On integration, we obtain
\begin{equation}\label{UFL_LOE:EQ3}
	\frac{\tilde{r}^{(n-1)\beta-(n+1)}_{\scaleto{A}{3.5pt}}}{(n-1)\beta-(n+1)}=-\frac{2\pi \Omega_{n}\rho}{\gamma \beta (n \Omega_{n})^{\beta}(n-1)}.
\end{equation}
Here, we have neglected the integration constant.
Multiply both sides of equation~\eqref{UFL_LOE:EQ3} by $a^2$ and then differentiate with respect to time. The resulting equation is then divided throughout by $2a\dot{a}$, and it leads to
\begin{equation}\label{UFL_LOE:EQ5}
	\left[(n-1)\beta-(n+1)\right]\tilde{r}^{(n-1)\beta-(n+2)}_{\scaleto{A}{3.5pt}}\frac{\dot{\tilde{r}}_{\scaleto{A}{3.5pt}}}{2H}+\tilde{r}^{(n-1)\beta-(n+1)}_{\scaleto{A}{3.5pt}}=\frac{\left[(n+1)-(n-1)\beta\right] 2\pi \Omega_{n}}{\gamma \beta (n \Omega_{n})^{\beta}(n-1)}\left(\frac{\dot{\rho}}{2H}+\rho\right).
\end{equation}
This equation can be further simplified by multiplying both sides with $4\alpha \gamma \beta (n \Omega_{n} \tilde{r}^{n-1}_{\scaleto{A}{3.5pt}})^{\beta}$ and  using continuity equation~\eqref{ce}, we thus get
\begin{equation}\label{UFL_LOE:EQ7}
	4\alpha \gamma  \frac{d\tilde{V}}{dt}=\tilde{r}_{\scaleto{A}{3.0pt}}H \left[\frac{8 \alpha \gamma \beta \tilde{A}}{\left[(n+1)-(n-1)\beta\right]}+\frac{4\pi \Omega_{n}\tilde{r}^{n+1}_{\scaleto{A}{3.5pt}}}{  n-2}\left((n-2)\rho+nP\right)\right],
\end{equation}
where $\tilde{A}=(n \Omega_{n} \tilde{r}^{n-1}_{\scaleto{A}{3.5pt}})^{\beta}$ is defined as the effective area of horizon, $\frac{d\tilde{V}}{dt}=\frac{\tilde{r}_{\scaleto{A}{3.0pt}}}{n-1}\frac{d\tilde{A}}{dt}$ is the rate of change of the corresponding effective volume of the horizon~\cite{SHEYKHI2018118, Chen2022}, and $\alpha=\frac{n-1}{2(n-2)}$. For $\beta=1$, the effective area $\tilde{A}$ and volume $\tilde{V}$ will reduce to area $A$ and the volume $V$.
The above equation~\eqref{UFL_LOE:EQ7} can be recast as the law of emergence
\begin{equation}\label{UFL_LOE:EQ8}
	4 \alpha\gamma  \frac{d\tilde{V}}{dt}=\tilde{r}_{\scaleto{A}{3.0pt}}H\left(N_{\text{sur}}-N_{\text{bulk}}\right),
\end{equation}
by identifying the degrees of freedom at the surface as $N_{\text{sur}}=\frac{\alpha 8\gamma \beta \tilde{A}}{\left[(n+1)-(n-1)\beta\right]}$ and that in the bulk as $N_{\text{bulk}}=-\frac{4\pi \Omega_{n}\tilde{r}^{n+1}_{\scaleto{A}{3.5pt}}}{  n-2}\left((n-2)\rho+nP\right)$, respectively. Since the surface degrees of freedom is positive, we will have the upper bound for the parameter $\beta$ as $(n+1)-(n-1)\beta>0$, that is, $\beta< \frac{n+1}{n-1}$. The bulk degrees of freedom have already emerged along with the space from some pre-geometrical variables~\cite{padmanabhan2012emergence}.
Equation~\eqref{UFL_LOE:EQ8} relates the emergence of space to the difference between the degrees of freedom in the surface and the bulk. This equation is identical to the law of emergence of cosmic space proposed by Chen~\cite{Chen2022} with Tsallis entropy. Here, in equation~\eqref{UFL_LOE:EQ8} the emergence of space is expressed in terms of the rate of change of effective volume. The modification in the surface degrees of freedom and the volume element is due to Tsallis entropy. From the equilibrium thermodynamic perspective, if we try to obtain the law of emergence in terms of the areal volume as follows:
equation~\eqref{UFL_LOE:EQ5} can be further simplified by applying continuity equation~\eqref{ce} and multiplying throughout by $\alpha n\Omega_{n}\tilde{r}^{n-1}_{\scaleto{A}{3.5pt}}$, then we get
\begin{equation}
	4\alpha \gamma  \frac{dV}{dt}=\tilde{r}_{\scaleto{A}{3.0pt}}H \left[\frac{8 \alpha \gamma \beta A}{\left[(n+1)-(n-1)\beta\right]}+\frac{4\pi \Omega_{n}\tilde{r}^{2n-(n-1)\beta}_{\scaleto{A}{3.5pt}}}{  \beta (n \Omega_{n})^{\beta-1}(n-2)}\left((n-2)\rho+nP\right)\right].
\end{equation}
To obtain the law of emergence in terms of areal volume, we must identify the terms on the right-hand side of the above equation as the surface and bulk degrees of freedom. That will result in $N_{\text{sur}}=\frac{8 \alpha \gamma \beta A}{\left[(n+1)-(n-1)\beta\right]}$ and $N_{\text{bulk}}=-\frac{4\pi \Omega_{n}\tilde{r}^{2n-(n-1)\beta}_{\scaleto{A}{3.5pt}}}{  \beta (n \Omega_{n})^{\beta-1}(n-2)}\left((n-2)\rho+nP\right)$. Hence, from the equilibrium thermodynamic perspective, it is clear that the law of emergence in terms of areal volume, with bulk degrees of freedom as defined in the existing literature, cannot be obtained.
For $n=3$, the equation~\eqref{UFL_LOE:EQ8} reduces to
\begin{equation}\label{loe-3+1}
4\gamma  \frac{d\tilde{V}}{dt}=\tilde{r}_{\scaleto{A}{3.0pt}}H\left(N_{\text{sur}}-N_{\text{bulk}}\right),
\end{equation}
where $N_{\text{sur}}=\frac{ 4\gamma \beta \tilde{A}}{2-\beta}$ and $N_{\text{bulk}}=-\frac{16\pi^{2}\tilde{r}^{4}_{\scaleto{A}{3.5pt}}}{ 3}\left(\rho+3P\right)$. This equation is identical to the law of emergence of cosmic space proposed by Sheykhi~\cite{SHEYKHI2018118} for $(3+1)$-dimensional space-time using Tsallis entropy.
For $\beta=1$ and $\gamma=\frac{1}{4L^{n-1}_{\scaleto{P}{3.5pt}}}$, at which the Tsallis entropy reduces to Bekenstein entropy, equation~\eqref{UFL_LOE:EQ8} reduces to 
\begin{equation}\label{loe-n+1}
\alpha \frac{d{V}}{dt}=L^{n-1}_{\scaleto{P}{3.5pt}}\tilde{r}_{\scaleto{A}{3.0pt}}H\left(N_{\text{sur}}-N_{\text{bulk}}\right),	\end{equation}
where $N_{\text{sur}}=\frac{\alpha A}{L^{n-1}_{\scaleto{P}{3.5pt}}}$ and $N_{\text{bulk}}=-\frac{4\pi \Omega_{n}\tilde{r}^{n+1}_{\scaleto{A}{3.5pt}}}{  n-2}\left((n-2)\rho+nP\right)$. This is similar to the law of emergence proposed by Sheykhi~\cite{PhysRevD.87.061501} in the $(n+1)$-dimensional non-flat FRW universe in the context of Einstein's gravity. Moreover, for a $(3+1)$-dimensional flat universe, the equation~\eqref{UFL_LOE:EQ8} reduces to
\begin{equation}
\frac{dV}{dt}=L_p^2(N_{\text{sur}}-N_{\text{bulk}}),
\end{equation}
for $\beta=1.$
Here $N_{\text{sur}}=\frac{A}{L^{n-1}_{\scaleto{P}{3.5pt}}}$ and $N_{\text{bulk}}=-\frac{16\pi^{2}} {3H^{4}}\left(\rho+3P\right).$ The above equation is identical to the original law of emergence proposed by Padmanabhan~\cite{padmanabhan2012emergence} for $(3+1)$-dimensional flat FRW universe in Einstein's gravity.
\subsection{Emergence of cosmic space from Clausius relation}
We will obtain the law of emergence from the Clausius relation~\eqref{CR1}, which is another form of the first law of thermodynamics, with Tsallis entropy, in this section. 
Considering the amount of energy crossing the apparent horizon during the infinitesimal time interval, $dt$, the equation~\eqref{CR1} becomes~\eqref{CR2} and substituting~\eqref{EFD} into it, we obtain
\begin{equation}\label{CR_LOE:EQ1}
(n \Omega_{n}  \tilde{r}^{n-1}_{\scaleto{A}{3.5pt}})(\rho + P)H dt = \frac{1}{2 \pi  \tilde{r}_{\scaleto{A}{3.0pt}}} \gamma \beta (n \Omega_{n})^{\beta}(n-1)\tilde{r}^{(n-1)\beta-1}_{\scaleto{A}{3.5pt}}d\tilde{r}_{\scaleto{A}{3.0pt}}.
\end{equation}
Using the continuity equation~\eqref{ce}, and then on rearrangement, the equation~\eqref{CR_LOE:EQ1} reduces to
\begin{equation}\label{CR_LOE:EQ2}
\tilde{r}^{(n-1)\beta-(n+2)}_{\scaleto{A}{3.5pt}} d\tilde{r}_{\scaleto{A}{3.0pt}}=
-\frac{2 \pi \Omega_{n}  d\rho }{\gamma \beta (n \Omega_{n})^{\beta}(n-1)}.
\end{equation}
The above equation on integration, neglecting the integration constant, will result in equation~\eqref{UFL_LOE:EQ3}. Then by following the steps in subsection~\ref{sec:1A}, we can obtain the law of emergence given in equation ~\eqref{UFL_LOE:EQ8}. Hence, it is possible to obtain the same law of emergence of cosmic space from both the unified first law and the Clausius relation.

\section{Emergence of cosmic space in Tsallis modified gravity from non-equilibrium perspective}
\label{s3}
In the previous section, we derived the law of emergence with Tsallis entropy from the equilibrium thermodynamic perspective. However, the necessity of using a non-equilibrium thermodynamic approach has been pointed out in the context of Tsallis entropy. For instance, Asghari and Sheykhi~\cite{10.1093/mnras/stab2671} showed that when Tsallis entropy is used as the horizon entropy, the conservation law, $\nabla^{\mu} T_{\mu \nu}=0$, is satisfied only if one adopts a non-equilibrium entropy balance relation, $dS = \frac{\delta Q}{T} + d_{i}S$,  where $d_{i}S$ is the additional entropy produced inside the horizon due to irreversible processes. This motivates the use of non-equilibrium thermodynamic relations in analyzing the evolution of the universe with Tsallis entropy as the horizon entropy. In this section, we obtain the law of emergence with Tsallis entropy from the non-equilibrium thermodynamic relations. For this, we adopt a general formulation developed by Tian and Booth~\cite{PhysRevD.90.104042}, which the authors used to obtain the Friedmann equations in modified gravity theories from thermodynamics with non-equilibrium considerations. In order to implement the formulation, we first restructure the Tsallis entropy given in~\eqref{ahs} into a form similar to Bekenstein-Hawking entropy, that is,
\begin{equation}\label{TE_NEQ}
S_{\scaleto{A}{3.5pt}}=\frac{A}{4G_{\scaleto{\text{eff}}{3.5pt}}}.
\end{equation}
Here, $\gamma$ in~\eqref{ahs} is suitably defined as $\gamma=\frac{(n\Omega_{n})^{1-\beta}}{4G\beta}$ and $G_{\scaleto{\text{eff}}{3.5pt}}=G\beta \tilde{r}^{n-1-(n-1)\beta}_{\scaleto{A}{3.5pt}}.$  
In the $(3+1)$ dimensions, $\gamma$ reduces to the form $\gamma=\frac{(4\pi)^{1-\beta}}{4G\beta}$ similar to that assumed by Asghari and Sheykhi~\cite{10.1093/mnras/stab2671}. The field equation for any modified gravity theory with horizon entropy as given in~\eqref{TE_NEQ} can be recast into a concise field equation similar to that in general relativity~\cite{PhysRevD.90.104042,PhysRevD.92.024001,M.Sharif_2012,ZUBAIR2016116},
\begin{equation}\label{FE_NEQ}
G_{\mu \nu}=8\pi G_{\scaleto{\text{eff}}{3.5pt}}T^{\text{(eff)}}_{\mu \nu}.
\end{equation}
Here $G_{\scaleto{\text{eff}}{3.5pt}}=G\beta \tilde{r}^{n-1-(n-1)\beta}_{\scaleto{A}{3.5pt}}$,  which is the same as in the equation~\eqref{TE_NEQ}, can be taken as an effective gravitational coupling. Unlike the field equation in Einstein's gravity, the effective gravitational coupling in the present case is time-dependent. However, the dependence on time differs from the idea of the time-varying gravitational constant over cosmic time. The effective energy-momentum tensor, $T^{\text{(eff)}}_{\mu \nu}=T^{\text{(m)}}_{\mu \nu}+T^{\text{(TG)}}_{\mu \nu}$, where $T^{\text{(m)}}_{\mu \nu}$ is the energy-momentum tensor of the physical matter content, and $T^{\text{(TG)}}_{\mu \nu}$ arises due to the Tsallis entropy. If we assume the cosmic component of the universe to be of perfect fluid type, then $T^{\mu\,\text{(eff)}}_{\nu}=(\rho_{\scaleto{\text{eff}}{3.5pt}}+P_{\scaleto{\text{eff}}{3.5pt}})u^{\mu}u_{\nu}+\delta^{\mu}_{\nu}P_{\scaleto{\text{eff}}{3.5pt}}$, where $\rho_{\scaleto{\text{eff}}{3.5pt}}=\rho_{\scaleto{m}{2.5pt}}+\rho_{\scaleto{\text{TG}}{3.0pt}}$ and $P_{\scaleto{\text{eff}}{3.5pt}}=P_{\scaleto{\text{m}}{3.5pt}}+P_{\scaleto{\text{TG}}{3.0pt}}$.
Here $\rho_{\scaleto{m}{2.5pt}}$ and $P_{\scaleto{m}{2.5pt}}$ are the energy density and pressure of all matter components in the universe, and $\rho_{\scaleto{\text{TG}}{3.0pt}}$ and $P_{\scaleto{\text{TG}}{3.0pt}}$ are the additional density and pressure arising due to the use of Tsallis entropy for the horizon. 
From the contracted Bianchi identity, the covariant derivative of the field equation~\eqref{FE_NEQ} satisfies, $\nabla_{\mu}G^{\mu}_{\nu}=0=\nabla_{\mu}(G_{\scaleto{\text{eff}}{3.5pt}}T^{\mu\,\text{(eff)}}_{\nu})$. This implies a generalized continuity equation~\cite{PhysRevD.90.104042}
\begin{equation}\label{CE_NEQ}
\dot{\rho}_{\scaleto{\text{eff}}{3.5pt}}+nH(\rho_{\scaleto{\text{eff}}{3.5pt}}+P_{\scaleto{\text{eff}}{3.5pt}})=-\frac{\dot{G_{\scaleto{\text{eff}}{3.5pt}}}}{G_{\scaleto{\text{eff}}{3.5pt}}}\rho_{\scaleto{\text{eff}}{3.5pt}}.
\end{equation} 
The non-zero term on the right-hand side of the above equation balances the energy flow, which has the dimension of the effective density flow.  The corresponding non-equilibrium energy dissipation term is $\mathcal{E}=-\Omega_{n}\tilde{r}^{n}_{\scaleto{A}{3.5pt}}\frac{\dot{G_{\scaleto{\text{eff}}{3.5pt}}}}{G_{\scaleto{\text{eff}}{3.5pt}}}\rho_{\scaleto{\text{eff}}{3.0pt}}dt$~\cite{PhysRevD.90.104042}.
Following this, the unified first law~\cite{Hayward_1998} modifies to~\cite{PhysRevD.90.104042}
\begin{equation}\label{UFL_NEQ}
dE_{\scaleto{\text{eff}}{3.5pt}}=A\psi_{\scaleto{\text{eff}}{3.5pt}}+W_{\scaleto{\text{eff}}{3.5pt}}dV+\mathcal{E},
\end{equation}
where $W_{\scaleto{\text{eff}}{3.5pt}}=\frac{\rho_{\scaleto{\text{eff}}{3.5pt}}- P_{\scaleto{\text{eff}}{3.5pt}}}{2}$ is the effective work density, and $E_{\scaleto{\text{eff}}{3.5pt}}=\rho_{\scaleto{\text{eff}}{3.5pt}} V$ is the effective total energy content inside the $n$ sphere of radius $\tilde{r}_{\scaleto{A}{3.0pt}}$. The effective energy flux density, in the context of the effective density and pressure, is 
\begin{align}\label{EFD_NEQ}
\psi_{\scaleto{\text{eff}}{3.5pt}}&=\psi_{t}^{\scaleto{\text{eff}}{3.5pt}}dt+\psi_{\tilde{r}_{\scaleto{A}{2.5pt}}}^{\scaleto{\text{eff}}{3.5pt}}d\tilde{r}_{\scaleto{A}{3.0pt}} 
=-(\rho_{\scaleto{\text{eff}}{3.5pt}}+P_{\scaleto{\text{eff}}{3.5pt}})H\tilde{r}_{\scaleto{A}{3.0pt}}dt+\frac{1}{2}(\rho_{\scaleto{\text{eff}}{3.5pt}}+P_{\scaleto{\text{eff}}{3.5pt}})d\tilde{r}_{\scaleto{A}{3.0pt}}.
\end{align}
Substituting the expressions for $\psi_{\scaleto{\text{eff}}{3.5pt}}$~\eqref{EFD_NEQ} and $W_{\scaleto{\text{eff}}{3.5pt}}$ in the unified first law~\eqref{UFL_NEQ} results in
\begin{equation}\label{UFL_NEQ_2}
dE_{\scaleto{\text{eff}}{3.5pt}}=A\psi_{t}^{\scaleto{\text{eff}}{3.5pt}}dt+A\psi_{\tilde{r}_{\scaleto{A}{3.0pt}}}^{\scaleto{\text{eff}}{3.5pt}}d\tilde{r}_{\scaleto{A}{3.0pt}} +\frac{\rho_{\scaleto{\text{eff}}{3.5pt}}- P_{\scaleto{\text{eff}}{3.5pt}}}{2}dV+\mathcal{E}
\end{equation}
Here, $(A\psi_{t}^{\scaleto{\text{eff}}{3.5pt}}dt+\mathcal{E})$ together constitute the energy crossing the horizon within an infinitesimal interval of time, $dt$, during which $\tilde{r}_{\scaleto{A}{3.0pt}}$ can be a constant. Then following the Clausius relation, the above energy flow is the heat flow $\delta Q$ through the apparent horizon within an infinitesimal time interval $dt$. As we mentioned, the term $\mathcal{E}$  corresponds to the non-equilibrium process, which will generate an additional entropy, $d_{\scaleto{P}{3.5pt}}S_{\scaleto{A}{3.5pt}}$. With these considerations, we can write the Clausius relation as~\cite{PhysRevD.90.104042}
\begin{equation}\label{CR_NEQ_1}
\delta Q=-A\psi_{t}^{\scaleto{\text{eff}}{3.5pt}}dt-\mathcal{E}=\tilde{T}_{\scaleto{A}{3.5pt}}(dS_{\scaleto{A}{3.5pt}}+d_{\scaleto{P}{3.5pt}}S_{\scaleto{A}{3.5pt}}).
\end{equation}
This can be considered as the non-equilibrium extension of the Clausius relation~\cite{PhysRevD.90.104042} with an irreversible extra entropy production term $d_{\scaleto{P}{3.5pt}}S_{\scaleto{A}{3.5pt}}$. 
These results simplify the non-equilibrium description of the unified first law~\eqref{UFL_NEQ_2} to the form~\cite{PhysRevD.92.024001}
\begin{align}\label{UFL_NEQ1}
dE_{\scaleto{\text{eff}}{3.5pt}}=-\frac{1}{2\pi \tilde{r}_{\scaleto{A}{3.0pt}}}(dS_{\scaleto{A}{3.5pt}}+d_{\scaleto{P}{3.5pt}}S_{\scaleto{A}{3.5pt}})+\rho_{\scaleto{\text{eff}}{3.5pt}}dV.
\end{align}
Multiply throughout the above equation~\eqref{UFL_NEQ1} with $\left( 1-\frac{ \dot{\tilde{r}}_{\scaleto{A}{3.5pt}}}{2H\tilde{r}_{\scaleto{A}{3.0pt}}} \right)$, and on further simplification, we arrive at another form of the unified first law given by
\begin{align}
dE_{\scaleto{\text{eff}}{3.5pt}}&=-T_{\scaleto{A}{3.5pt}}(dS_{\scaleto{A}{3.5pt}}+d_{\scaleto{P}{3.5pt}}S_{\scaleto{A}{3.5pt}})+W_{\scaleto{\text{eff}}{3.5pt}}dV+ \frac{\mathcal{E}\dot{\tilde{r}}_{\scaleto{A}{3.5pt}}}{2H\tilde{r}_{\scaleto{A}{3.0pt}}}. \label{UFL_NEQ2}
\end{align}
To proceed further, we have to extract the entropy production term $d_{\scaleto{P}{3.5pt}}S_{\scaleto{A}{3.5pt}}$. 
By rearranging the unified first law~\eqref{UFL_NEQ2}, the extra entropy production term can be written as
\begin{equation}\label{EP_NEQ_EQ1}
T_{\scaleto{A}{3.5pt}}d_{\scaleto{P}{3.5pt}}S_{\scaleto{A}{3.5pt}}=-T_{\scaleto{A}{3.5pt}}dS_{\scaleto{A}{3.5pt}}+W_{\scaleto{\text{eff}}{3.5pt}}dV+ \frac{\mathcal{E}\dot{\tilde{r}}_{\scaleto{A}{3.5pt}}}{2H\tilde{r}_{\scaleto{A}{3.0pt}}}-dE_{\scaleto{\text{eff}}{3.5pt}}.
\end{equation}
Substituting $T_{\scaleto{A}{3.5pt}}$, $S_{\scaleto{A}{3.5pt}}$, $W_{\scaleto{\text{eff}}{3.5pt}}$, $E_{\scaleto{\text{eff}}{3.5pt}}$, and $V$ in the above expression and on
simplification we get
\begin{align}\label{EP_NEQ_EQ3_1}
T_{\scaleto{A}{3.5pt}}d_{\scaleto{P}{3.5pt}}S_{\scaleto{A}{3.5pt}}=&\left[1-\frac{\dot{\tilde{r}}_{\scaleto{A}{3.5pt}}}{2H\tilde{r}_{\scaleto{A}{3.0pt}}}\right]\left[\frac{n \Omega_{n}\tilde{r}^{n-2}_{\scaleto{A}{3.5pt}}\dot{G_{\scaleto{\text{eff}}{3.5pt}}}}{8 \pi G^{2}_{\scaleto{\text{eff}}{3.5pt}}}-\frac{n(n-1)\Omega_{n}\tilde{r}^{n-3}_{\scaleto{A}{3.5pt}}\dot{\tilde{r}}_{\scaleto{A}{3.5pt}}}{8 \pi G_{\scaleto{\text{eff}}{3.5pt}}}\right]dt+\frac{-n\Omega_{n}\tilde{r}^{n-1}_{\scaleto{A}{3.5pt}}\dot{\tilde{r}}_{\scaleto{A}{3.5pt}}dt}{2}\left[(\rho_{\scaleto{\text{eff}}{3.5pt}}+P_{\scaleto{\text{eff}}{3.5pt}})+\frac{\dot{G_{\scaleto{\text{eff}}{3.5pt}}}\rho_{\scaleto{\text{eff}}{3.5pt}}}{nHG_{\scaleto{\text{eff}}{3.5pt}}}\right]\nonumber \\
&-\Omega_{n}\tilde{r}^{n}_{\scaleto{A}{3.5pt}}d\rho_{\scaleto{\text{eff}}{3.5pt}}.
\end{align}
Applying the continuity equation~\eqref{CE_NEQ} will reduce the above equation to
\begin{align}\label{EP_NEQ_EQ3_2}
T_{\scaleto{A}{3.5pt}}d_{\scaleto{P}{3.5pt}}S_{\scaleto{A}{3.5pt}}=&\left[1-\frac{\dot{\tilde{r}}_{\scaleto{A}{3.5pt}}}{2H\tilde{r}_{\scaleto{A}{3.0pt}}}\right]\left[\frac{n \Omega_{n}\tilde{r}^{n-2}_{\scaleto{A}{3.5pt}}\dot{G_{\scaleto{\text{eff}}{3.5pt}}}}{8 \pi G^{2}_{\scaleto{\text{eff}}{3.5pt}}}-\frac{n(n-1)\Omega_{n}\tilde{r}^{n-3}_{\scaleto{A}{3.5pt}}\dot{\tilde{r}}_{\scaleto{A}{3.5pt}}}{8 \pi G_{\scaleto{\text{eff}}{3.5pt}}}+n \Omega_{n} \tilde{r}^{n}_{\scaleto{A}{3.5pt}}H(\rho_{\scaleto{\text{eff}}{3.5pt}}+P_{\scaleto{\text{eff}}{3.5pt}})+\frac{\Omega_{n}\tilde{r}^{n}_{\scaleto{A}{3.5pt}}\dot{G_{\scaleto{\text{eff}}{3.5pt}}}\rho_{\scaleto{\text{eff}}{3.5pt}}}{G_{\scaleto{\text{eff}}{3.5pt}}}\right]dt.
\end{align}
This can be further simplified by substituting $\rho_{\scaleto{\text{eff}}{3.5pt}}$ and $P_{\scaleto{\text{eff}}{3.5pt}}$ using the Friedmann equations. The Friedmann equations can be obtained by substituting the FRW metric~\eqref{n+1m} and the effective energy-momentum tensor into the field equation~\eqref{FE_NEQ} and are~\cite{PhysRevD.90.104042}
\begin{equation}\label{FLRW1_NEQ}
H^{2}+\frac{k}{a^2}=\frac{16\pi G_{\scaleto{\text{eff}}{3.5pt}}}{n(n-1)}\rho_{\scaleto{\text{eff}}{3.5pt}}\quad \text{and}
\end{equation}
\begin{equation}\label{FLRW2_NEQ}
\dot{H}-\frac{k}{a^2}=-\frac{8\pi G_{\scaleto{\text{eff}}{3.5pt}}}{(n-1)}(\rho_{\scaleto{\text{eff}}{3.5pt}}+P_{\scaleto{\text{eff}}{3.5pt}}).
\end{equation}
From the equations~\eqref{EP_NEQ_EQ3_2},~\eqref{FLRW1_NEQ}, and~\eqref{FLRW2_NEQ}, we have
\begin{equation}\label{EP_NEQ_EQ4}
T_{\scaleto{A}{3.5pt}}d_{\scaleto{P}{3.5pt}}S_{\scaleto{A}{3.5pt}}=\left[1-\frac{\dot{\tilde{r}}_{\scaleto{A}{3.5pt}}}{2H\tilde{r}_{\scaleto{A}{3.0pt}}}\right]\frac{n(n+1) \Omega_{n}\tilde{r}^{n-2}_{\scaleto{A}{3.5pt}}\dot{G_{\scaleto{\text{eff}}{3.5pt}}}}{16\pi G^{2}_{\scaleto{\text{eff}}{3.5pt}}}dt.
\end{equation}
After substituting for the temperature~\eqref{aht} in the above expression, we will obtain the entropy production term as
\begin{equation}\label{EP_NEQ_EQ5}	
d_{\scaleto{P}{3.5pt}}S_{\scaleto{A}{3.5pt}}=\frac{n(n+1) \Omega_{n}\tilde{r}^{n-1}_{\scaleto{A}{3.5pt}}\dot{G_{\scaleto{\text{eff}}{3.5pt}}}}{8G^{2}_{\scaleto{\text{eff}}{3.5pt}}}dt.
\end{equation}
By adopting a similar procedure it can be shown that the extra entropy production term can also be calculated from the non-equilibrium extension of the Clausius relation~\eqref{CR_NEQ_1}. We will obtain the law of emergence in the following subsection with these results.
\subsection{Emergence of cosmic space from the non-equilibrium description of the unified first law of thermodynamics}\label{sec:2A}
In this section, we will consider the non-equilibrium description of the unified first law of thermodynamics~\eqref{UFL_NEQ2} to derive the law of emergence. Substituting the expressions for the temperature $T_{\scaleto{A}{3.5pt}}$~\eqref{aht}, the horizon entropy $S_{\scaleto{A}{3.5pt}}$~\eqref{TE_NEQ}, the entropy production term $d_{\scaleto{P}{3.5pt}}S_{\scaleto{A}{3.5pt}}$~\eqref{EP_NEQ_EQ5}, the effective energy $E_{\scaleto{\text{eff}}{3.5pt}}=\rho_{\scaleto{\text{eff}}{3.0pt}} V$, the effective work density $W_{\scaleto{\text{eff}}{3.5pt}}$, and the non-equilibrium energy dissipation term $\mathcal{E}$ into the unified first law~\eqref{UFL_NEQ2} results in 
\begin{equation}\label{EP_NEQ_EQ6}
\begin{aligned}
	&\left[\frac{1}{2}(\rho_{\scaleto{\text{eff}}{3.5pt}}+P_{\scaleto{\text{eff}}{3.5pt}})-\frac{\dot{G_{\scaleto{\text{eff}}{3.5pt}}}\rho_{\scaleto{\text{eff}}{3.5pt}}}{2nHG_{\scaleto{\text{eff}}{3.5pt}}}\right]dV+Vd\rho_{\scaleto{\text{eff}}{3.5pt}}=-\left[1-\frac{\dot{\tilde{r}}_{\scaleto{A}{3.5pt}}}{2H\tilde{r}_{\scaleto{A}{3.0pt}}}\right]\left[\frac{n(n-1)
		\Omega_{n}\tilde{r}^{n-2}_{\scaleto{A}{3.5pt}}\dot{G_{\scaleto{\text{eff}}{3.5pt}}}}{16 \pi G^{2}_{\scaleto{\text{eff}}{3.5pt}}}\right. \left.+\frac{n(n-1)\Omega_{n}\tilde{r}^{n-3}_{\scaleto{A}{3.5pt}}\dot{\tilde{r}}_{\scaleto{A}{3.5pt}}}{8 \pi G_{\scaleto{\text{eff}}{3.5pt}}}\right]dt
\end{aligned}
\end{equation}
Applying the continuity equation~\eqref{CE_NEQ} and the Friedmann equation~\eqref{FLRW1_NEQ}, reduces the above equation to 
\begin{equation}\label{EP_NEQ_EQ7}
-2\frac{\dot{\tilde{r}}_{\scaleto{A}{3.5pt}}dt}{\tilde{r}^{3}_{\scaleto{A}{3.5pt}}}=\frac{16\pi(\dot{G_{\scaleto{\text{eff}}{3.5pt}}}\rho_{\scaleto{\text{eff}}{3.5pt}}+G_{\scaleto{\text{eff}}{3.5pt}}\dot{\rho}_{\scaleto{\text{eff}}{3.5pt}})dt}{n(n-1)}.
\end{equation} 
On integrating the above expression, we obtain
\begin{equation}\label{EP_NEQ_EQ8}
\frac{1}{\tilde{r}^{2}_{\scaleto{A}{3.5pt}}}=\frac{16\pi G_{\scaleto{\text{eff}}{3.5pt}}\rho_{\scaleto{\text{eff}}{3.5pt}}}{n(n-1)}.
\end{equation}
Here we have neglected the integration constant for convenience. Multiply both sides of~\eqref{EP_NEQ_EQ8} by $a^2$ and differentiate with respect to time on both sides. Then dividing both sides of the resultant equation by  $2a\dot{a}$, we get
\begin{equation}\label{EP_NEQ_EQ10}
-\frac{\dot{\tilde{r}}_{\scaleto{A}{3.5pt}}}{\tilde{r}^{3}_{\scaleto{A}{3.5pt}}H}+\frac{1}{\tilde{r}^{2}_{\scaleto{A}{3.5pt}}}=\frac{16\pi}{n(n-1)}\left[\frac{\dot{G_{\scaleto{\text{eff}}{3.5pt}}}\rho_{\scaleto{\text{eff}}{3.5pt}}+G_{\scaleto{\text{eff}}{3.5pt}}\dot{\rho}_{\scaleto{\text{eff}}{3.5pt}}}{2H}+G_{\scaleto{\text{eff}}{3.5pt}}\rho_{\scaleto{\text{eff}}{3.5pt}}\right].
\end{equation}
Rearranging the equation and using the continuity equation will result in
\begin{equation}\label{EP_NEQ_EQ11}
\dot{\tilde{r}}_{\scaleto{A}{3.5pt}}=\tilde{r}_{\scaleto{A}{3.0pt}}H-\frac{16\pi\tilde{r}^{3}_{\scaleto{A}{3.5pt}}H}{n(n-1)}\left[-\frac{nG_{\scaleto{\text{eff}}{3.5pt}}(\rho_{\scaleto{\text{eff}}{3.5pt}}+P_{\scaleto{\text{eff}}{3.5pt}})}{2}+G_{\scaleto{\text{eff}}{3.5pt}}\rho_{\scaleto{\text{eff}}{3.5pt}}\right].
\end{equation}
Multiplying~\eqref{EP_NEQ_EQ11} by $\alpha n\Omega_{n}\tilde{r}^{n-1}_{\scaleto{A}{3.5pt}}$ can reduce the expression to
\begin{equation}\label{EP_NEQ_EQ12}
\alpha n\Omega_{n}\tilde{r}^{n-1}_{\scaleto{A}{3.5pt}}\dot{\tilde{r}}_{\scaleto{A}{3.5pt}}=\alpha n\Omega_{n}\tilde{r}^{n}_{\scaleto{A}{3.5pt}}H+\frac{ 4\pi \Omega_{n}\tilde{r}^{n+2}_{\scaleto{A}{3.5pt}}G_{\scaleto{\text{eff}}{3.5pt}}H}{(n-2)}\left[(n-2)\rho_{\scaleto{\text{eff}}{3.5pt}}+nP_{\scaleto{\text{eff}}{3.5pt}}\right].
\end{equation}
The above expression can be written as
\begin{equation}\label{EP_NEQ_EQ13}
\alpha \frac{dV}{dt}=\tilde{r}_{\scaleto{A}{3.0pt}}HG_{\scaleto{\text{eff}}{3.5pt}}\left[\frac{\alpha A}{G_{\scaleto{\text{eff}}{3.5pt}}}+\frac{ 4\pi \Omega_{n}\tilde{r}^{n+1}_{\scaleto{A}{3.5pt}}}{(n-2)}\left[(n-2)\rho_{\scaleto{\text{eff}}{3.5pt}}+nP_{\scaleto{\text{eff}}{3.5pt}}\right]\right].
\end{equation}
This equation can be identified as the
\begin{equation}\label{EP_NEQ_EQ14}
\alpha \frac{dV}{dt}=\tilde{r}_{\scaleto{A}{3.0pt}}HG_{\scaleto{\text{eff}}{3.5pt}}\left(N_{\text{sur}}-N_{\text{bulk}}\right),
\end{equation}
where $N_{\text{sur}}=\frac{\alpha A}{G_{\scaleto{\text{eff}}{3.5pt}}}$ and $N_{\text{bulk}}=-\frac{ 4\pi \Omega_{n}\tilde{r}^{n+1}_{\scaleto{A}{3.5pt}}}{(n-2)}\left[(n-2)\rho_{\scaleto{\text{eff}}{3.5pt}}+nP_{\scaleto{\text{eff}}{3.5pt}}\right]$.  
For $\beta=1$, $G_{\scaleto{\text{eff}}{3.5pt}}$, $\rho_{\scaleto{\text{eff}}{3.5pt}}$ and $P_{\scaleto{\text{eff}}{3.5pt}}$ reduces to $G$, $\rho_{\scaleto{\text{m}}{3.5pt}}$ and $P_{\scaleto{\text{m}}{3.5pt}}$, consequently the equation~\eqref{EP_NEQ_EQ14} becomes identical to the law of emergence~\eqref{loe-n+1} proposed by Sheykhi~\cite{PhysRevD.87.061501} in the $(n+1)$-dimensional space-time in Einstein's gravity. The interesting feature regarding the above form of the law is that the volume appearing in the left hand side is areal volume of the apparent horizon. While in the corresponding form of the law in the equilibrium case contains an effective volume of the horizon. 
This simplicity in the appearance of the law of emergence in the non-equilibrium case is  
due to the shifting and consequent absorption of the modifications in the curvature part of field equation into the matter part (see appendix~\ref{TMG_FE}).
Another interesting result is that, the law of emergence given in~\eqref{EP_NEQ_EQ14} can be treated as the general formula for any modified gravity theories, provided the horizon entropy in it can written in a form $S=\frac{A}{4G_{\scaleto{\text{eff}}{3.5pt}}}$. 

\subsection{Emergence of cosmic space from non-equilibrium description of the Clausius relation}
In this section, we will show that the same form of the law of emergence can be obtained from the Clausius relation, an alternative form of the first law of thermodynamics. 
Substituting the expressions for the horizon entropy $S_{\scaleto{A}{3.5pt}}$~\eqref{TE_NEQ}, the entropy production term $d_{\scaleto{P}{3.5pt}}S_{\scaleto{A}{3.5pt}}$~\eqref{EP_NEQ_EQ5}, and the non-equilibrium energy dissipation term $\mathcal{E}$ into the non-equilibrium extension of the Clausius relation~\eqref{CR_NEQ_1} will result in 
\begin{equation}\label{NECL:EQ6}
n\Omega_{n}\tilde{r}^{n-1}_{\scaleto{A}{3.5pt}}(\rho_{\scaleto{\text{eff}}{3.5pt}}+P_{\scaleto{\text{eff}}{3.5pt}})H\tilde{r}_{\scaleto{A}{3.0pt}}dt+\Omega_{n}\tilde{r}^{n}_{\scaleto{A}{3.5pt}}\frac{\dot{G_{\scaleto{\text{eff}}{3.5pt}}}}{G_{\scaleto{\text{eff}}{3.5pt}}}\rho_{\scaleto{\text{eff}}{3.5pt}}dt=
\frac{1}{2\pi \tilde{r}_{\scaleto{A}{3.0pt}}} \left[\frac{n(n-1)\Omega_{n}\tilde{r}^{n-2}_{\scaleto{A}{3.5pt}}\dot{\tilde{r}}_{\scaleto{A}{3.5pt}}}{4G_{\scaleto{\text{eff}}{3.5pt}}}+\frac{n(n-1) \Omega_{n}\tilde{r}^{n-1}_{\scaleto{A}{3.5pt}}\dot{G_{\scaleto{\text{eff}}{3.5pt}}}}{8G^{2}_{\scaleto{\text{eff}}{3.5pt}}}\right]dt.
\end{equation}
Using the continuity equation~\eqref{CE_NEQ} and the Friedmann equation~\eqref{FLRW1_NEQ}, the above equation can be reduced to the equation~\eqref{EP_NEQ_EQ7}. Then by following the steps mentioned in section~\ref{sec:2A}, we can obtain the law of emergence of cosmic space in Tsallis modified gravity. Hence, it is possible to obtain the same law of emergence in Tsallis modified gravity from both the non-equilibrium unified first law and the Clausius relation. In the non-equilibrium treatment, we started with rewriting the Tsallis entropy in the Bekenstein-Hawking entropy form~\eqref{TE_NEQ} with $G_{\scaleto{\text{eff}}{3.5pt}}=G\beta \tilde{r}^{n-1-(n-1)\beta}_{\scaleto{A}{3.5pt}}$ by defining $\gamma=\frac{(n\Omega_{n})^{1-\beta}}{4G\beta}$. Even if the $\gamma$ is not in the simple specified form, the above formulation will be valid since it is possible to define $G_{\scaleto{\text{eff}}{3.5pt}}$ such that Tsallis entropy takes the form~\eqref{TE_NEQ}.

\subsection{Emergence of cosmic space and maximization of entropy from the non-equilibrium perspective}

Any isolated macroscopic systems will evolve towards a state of maximum entropy~\cite{Callen:1960:T}, subject to the constraints, which implies that (i) the total entropy of an isolated system cannot decrease, that is, the generalized second law, $\dot{S}\geq0,$  and (ii) the entropy should be a convex function, that is, $\ddot{S}<0$, at least in the final stage of evolution~\cite{Pavon2013}. In the context of Einstein's gravity, it is shown that the entropy of our universe tends to a maximum~\cite{Pavon2013}. This tendency of the universe to achieve the state of maximum entropy  has also been shown in the context of  the law of emergence~\cite{PhysRevD.96.063513}. According to the law of emergence, the universe expands to achieve the state of 
holographic equipartition, $N_{\text{sur}}=N_{\text{bulk}}$, at which the universe enters a de Sitter epoch, corresponding to which the entropy becomes maximum. This result has been extended to 
Gauss-Bonnet and Lovelock gravity~\cite{PhysRevD.99.023535} theories.  It is to be noted that these results are in the framework of equilibrium thermal evolution. Meantime, the consistency of the law of emergence, 
with the maximization of entropy in the context of Tsallis entropy, has also been proven~\cite{Chen2022}, but in the equilibrium context. This motivates us to check the consistency of the law of emergence, obtained from the non-equilibrium perspective~\eqref{EP_NEQ_EQ14} in the context of Tsallis entropy, with the entropy maximization principle. 
Here we obtain the constraints
to satisfy the principle of entropy maximization. Further, we check the feasibility of these constraints in an expanding universe, which is approaching a final de Sitter epoch. The entropy mentioned here is the total entropy comprising the entropy of the  matter content of the universe and the horizon entropy. Since we consider Tsallis entropy as the horizon entropy, we will also have to incorporate the extra non-equilibrium entropy production. In this paper, we neglect the entropy of the  matter content since the matter entropy is much less than the horizon entropy~\cite{Egan_2010,https://doi.org/10.48550/arxiv.2002.02121,PhysRevD.15.2738}. To obtain the constraints for entropy maximization, let us consider the left-hand side of the law of emergence~\eqref{EP_NEQ_EQ14}, the time derivative of cosmic volume in $(n+1)-$dimensional universe, which is given by
\begin{equation}\label{HEM4}
\frac{dV}{dt}=n\Omega_{n}\tilde{r}^{n-1}_{\scaleto{A}{3.5pt}}  \dot{\tilde{r}}_{\scaleto{A}{3.5pt}}.
\end{equation}
The rate of  change of total entropy using the equations~\eqref{TE_NEQ} and~\eqref{EP_NEQ_EQ5} is
\begin{equation}\label{HEM1}
\dot{S}=\dot{S}_{\scaleto{A}{3.5pt}}+d_{\scaleto{P}{3.5pt}}\dot{S}_{\scaleto{A}{3.5pt}}=\frac{n\Omega_{n}(n-1)\left(n+1-\beta\left(n-1\right)\right)}{16 G \beta}\tilde{r}^{(n-1)\beta-1}_{\scaleto{A}{3.5pt}}\dot{\tilde{r}}_{\scaleto{A}{3.5pt}}.
\end{equation}
Comparing the equations~\eqref{HEM4} and~\eqref{HEM1}, we can write
\begin{equation}\label{HEM5}
\frac{dV}{dt}=\frac{16 G \beta \tilde{r}^{n-(n-1)\beta}_{\scaleto{A}{3.5pt}}}{(n-1)\left(n+1-\beta\left(n-1\right)\right)}		\dot{S}_{\scaleto{A}{3.5pt}}+d_{\scaleto{P}{3.5pt}}\dot{S}_{\scaleto{A}{3.5pt}}.
\end{equation}
Using the law of emergence~\eqref{EP_NEQ_EQ14}, the equation~\eqref{HEM5} can be written as
\begin{equation}\label{HEM6}
\dot{S}= \frac{H(n-2)\left(n+1-\beta\left(n-1\right)\right)}{8}\left(N_{\text{sur}}-N_{\text{bulk}}\right).
\end{equation}
Here $N_{\text{sur}}-N_{\text{bulk}}\geq 0$, which implies that the total entropy is non-decreasing for $n+1-\beta\left(n-1\right)  > 0$. Hence, the generalized second law is satisfied for $\beta<\frac{n+1}{n-1}$. We assume the upper bound for the parameter $\beta$ obtained from the maximization condition to be true for the following.
Now to check the convexity condition, the second derivative of entropy is to be considered, which can be obtained from the above equation by differentiating with the cosmic time, and we get as,
\begin{align}\label{HEM7}
\ddot{S}&=\ddot{S}_{\scaleto{A}{3.5pt}}+d_{\scaleto{P}{3.5pt}}	\ddot{S}_{\scaleto{A}{3.5pt}}=\frac{\left(n+1-\beta\left(n-1\right)\right)}{8}
\left[(n-2)\dot{H}\left(N_{\text{sur}}-N_{\text{bulk}}\right)+(n-2)H\left(\dot{N}_{\text{sur}}-\dot{N}_{\text{bulk}}\right)\right].
\end{align}
In an asymptotically de Sitter universe, in the final stage, $N_{\text{sur}}=N_{\text{bulk}}$, hence the first term in the above equation will vanish and for $\ddot{S}<0$, $\dot{N}_{\text{sur}}-\dot{N}_{\text{bulk}}$ should be less than zero. 
From the definition of bulk and surface degrees of freedom in equation~\eqref{EP_NEQ_EQ14} and using the Friedmann equations~\eqref{FLRW1_NEQ} and~\eqref{FLRW2_NEQ}, we know that
\begin{equation}\label{HEM9}
N_{\text{sur}}-N_{\text{bulk}}=\frac{ n(n-1) \Omega_{n}\tilde{r}^{(n-1)\beta-1}_{\scaleto{A}{3.5pt}}\dot{\tilde{r}}_{\scaleto{A}{3.5pt}}}{2(n-2) GH \beta}.
\end{equation}
Substituting the above equation~\eqref{HEM9} and its derivative
in to~\eqref{HEM7} results to
\begin{equation}\label{HEM3}
\ddot{S}=\frac{n\Omega_{n}(n-1)\left(n+1-\beta\left(n-1\right)\right)}{16 G \beta}\tilde{r}^{(n-1)\beta-2}_{\scaleto{A}{3.5pt}}\left[((n-1)\beta -1)\dot{\tilde{r}}^{2}_{\scaleto{A}{3.5pt}}+\tilde{r}_{\scaleto{A}{3.0pt}}\ddot{\tilde{r}}_{\scaleto{A}{3.5pt}}\right].
\end{equation}
The derivative of equation~\eqref{HEM1} will also lead to the same result~\eqref{HEM3}. For $\beta=1$, the above  expressions of the first and second derivatives of total entropy~\eqref{HEM1} reduce to the corresponding derivatives of entropy in  Einstein's gravity~\cite{https://doi.org/10.48550/arxiv.2002.02121}. 
From the above equation, the condition for the second derivative of the entropy to be negative, at least in the final stage, is
\begin{equation}\label{HEM10}
((n-1)\beta -1)\dot{\tilde{r}}^{2}_{\scaleto{A}{3.5pt}}<-\tilde{r}_{\scaleto{A}{3.0pt}}\ddot{\tilde{r}}_{\scaleto{A}{3.5pt}}.
\end{equation}
To check the validity of the above condition, 
let us consider the relation of $\dot{\tilde{r}}_{\scaleto{A}{3.5pt}}$  in terms of the effective equation of state parameter $\omega_{\scaleto{\text{eff}}{3.5pt}}$, taken from the equation of state $P_{\scaleto{\text{eff}}{3.5pt}}=\omega_{\scaleto{\text{eff}}{3.5pt}} \rho_{\scaleto{\text{eff}}{3.5pt}}$, written using the Friedmann equations~\eqref{FLRW1_NEQ} and~\eqref{FLRW2_NEQ} and the continuity equation~\eqref{CE_NEQ}, in the following way
\begin{equation}\label{HEM2}
\dot{\tilde{r}}_{\scaleto{A}{3.5pt}}=\frac{H\tilde{r}_{\scaleto{A}{3.0pt}}(1+\omega_{\scaleto{\text{eff}}{3.5pt}})}{2}.
\end{equation}
Taking the derivative of the above equation will give a relation connecting $\ddot{\tilde{r}}_{\scaleto{A}{3.5pt}}$ and $\omega_{\scaleto{\text{eff}}{3.5pt}}$ as follows
\begin{equation}\label{HEM11}
\ddot{\tilde{r}}_{\scaleto{A}{3.5pt}}=\frac{1}{2}\left[\tilde{r}_{\scaleto{A}{3.0pt}}(1+\omega_{\scaleto{\text{eff}}{3.5pt}})\left[\dot{H}+\frac{H^{2}(1+\omega_{\scaleto{\text{eff}}{3.5pt}})}{2}\right]+H\tilde{r}_{\scaleto{A}{3.0pt}}\dot{\omega}_{\scaleto{\text{eff}}{3.5pt}}\right].
\end{equation}
For an asymptotically de Sitter universe, $\omega_{\scaleto{\text{eff}}{3.5pt}}$ tends to $-1$, then $\dot{\tilde{r}}_{\scaleto{A}{3.5pt}}=0$, which reduces the condition~\eqref{HEM10} to $-\tilde{r}_{\scaleto{A}{3.0pt}}\ddot{\tilde{r}}_{\scaleto{A}{3.5pt}}>0$. For $\ddot{S}<0$ in an asymptotically de Sitter universe, $\ddot{\tilde{r}}_{\scaleto{A}{3.5pt}}$ should be negative. The first term in the above equation vanishes, and the negativity of $\dot{\omega}_{\scaleto{\text{eff}}{3.5pt}}$ will guarantee $\ddot{\tilde{r}}_{\scaleto{A}{3.5pt}}<0$, thereby showing $\ddot{S}<0$ in the final stage. It is clear from the equations~\eqref{HEM1},~\eqref{HEM6}, and~\eqref{HEM9} that to satisfy the condition that the total entropy does not decrease, $\dot{\tilde{r}}_{\scaleto{A}{3.5pt}}$ should always be greater than or equal to zero. That indicates the consistency of the law of emergence with the generalized second law. The equation~\eqref{HEM2} also guarantees that  $\dot{\tilde{r}}_{\scaleto{A}{3.5pt}}\geq0$ for an asymptotically de Sitter universe with $\omega_{\scaleto{\text{eff}}{3.5pt}}\geq-1$.  Hence we obtained the constraints for entropy maximization from the law of emergence in the non-equilibrium context, and these constraints are satisfied for an asymptotically de Sitter universe. The upper bound for the Tsallis parameter $\beta<\frac{n+1}{n-1}$, obtained from the constraints imposed by the generalized second law and the entropy maximization principle on the law of emergence, is a result of the non-equilibrium thermodynamic approach. Whereas in the case of equilibrium thermodynamic approach, such an upper bound for the Tsallis parameter $\beta$ arises from the surface degrees of freedom~\eqref{UFL_LOE:EQ8}. In $(3+1)$-dimensional universe, according to the above constraint, $\beta<2$, and is found to be true from the observational constraints~\cite{10.1093/mnras/stab2671}.
\section{Conclusions}
The connection between gravity and thermodynamics leads to the idea that gravity could be an emergent phenomenon. Following this, Padmanabhan proposed that the space could also be an emergent structure. Based on this, the expansion of the universe can be thought of as the emergence of space with cosmic time. The explanation for the expansion of the universe in terms of the degrees of freedom on the boundary and the bulk is given by the law of emergence, $\frac{dV}{dt}=L_p^2(N_{\text{sur}}-N_{\text{bulk}})$. A deep connection exists between the law of emergence and the fundamental thermodynamic laws, as one can derive the former from the latter. In this paper, we explored this connection in the context of Tsallis entropy in both equilibrium and non-equilibrium perspectives. We have considered Tsallis entropy, $S_{\scaleto{A}{3.5pt}}=\gamma A^{\beta}$, as the entropy associated with the apparent horizon and derived the law of emergence from the fundamental thermodynamic laws. Firstly, from an equilibrium perspective, we derived the law of emergence from the unified first law of thermodynamics and also from the Clausius relation. The law of emergence thus obtained substantiates the proposal of Sheykhi~\cite{SHEYKHI2018118} and Chen~\cite{Chen2022}. 

In the case of Tsallis entropy, the equilibrium Clausius relation does not hold due to non-equilibrium thermal evolution, which resulted in the production of an additional entropy~\cite{10.1093/mnras/stab2671}. This motivated us to obtain the law of emergence from the non-equilibrium perspective. First, we have restructured the Tsallis entropy into a form $S_{\scaleto{A}{3.5pt}}=\frac{A}{4G_{\scaleto{\text{eff}}{3.5pt}}}$, similar to the standard form of the Bekenstein-Hawking entropy, by assuming $\gamma=\frac{(n\Omega_{n})^{1-\beta}}{4G\beta}$ and $G_{\scaleto{\text{eff}}{3.5pt}}=G\beta \tilde{r}^{(n-1)(1-\beta)}_{\scaleto{A}{3.5pt}}$. Consequently, the gravitational field equation of Tsallis modified gravity takes a concise form $G_{\mu \nu}=8\pi G_{\scaleto{\text{eff}}{3.5pt}}T^{\text{(eff)}}_{\mu \nu}$, similar to the Einstein's gravity. The covariant derivative of the field equation then gives the continuity equation with a non-zero term on the right-hand side~\eqref{CE_NEQ}. The corresponding non-equilibrium energy dissipation,  $\mathcal{E}$, generates an additional entropy, $d_{\scaleto{P}{3.5pt}}S_{\scaleto{A}{3.5pt}}$, which is evaluated using Friedmann equations. By incorporating these into the unified first law and the Clausius relation, we have obtained the law of emergence.

The law of emergence in the equilibrium case describes the evolution of the space as the one which emerges in proportion to the rate of change of the effective volume and not the areal volume. As mentioned in appendix~\ref{a1}, the Friedmann equation obtained from the two different equilibrium approaches has different assumptions for $\gamma$. These problems can be resolved by following the non-equilibrium approach discussed in section~\ref{s3}. Since we restructure the Tsallis entropy into the form~\eqref{TE_NEQ} in the non-equilibrium approach, the question of the form of $\gamma$ does not arise here because it is possible to define any $G_{\scaleto{\text{eff}}{3.5pt}}$ according to the form of $\gamma$ that does not affect the law of emergence~\eqref{EP_NEQ_EQ14} derived here. The law of emergence in the non-equilibrium perspective describes the volume increase due to the accelerated expansion of the universe as the rate of change of the areal volume. In the non-equilibrium approach, we have defined an effective energy-momentum tensor that holds the corrections due to the Tsallis entropy rather than keeping the corrections on the curvature part in the field equation. That is to formulate a general law of emergence, which is impossible in the latter case (see appendix~\ref{TMG_FE}). The lucidity in the form of the law of emergence in the non-equilibrium case is mainly due to the definition of the effective energy-momentum tensor, $T^{\text{(eff)}}_{\mu \nu}$. It can be concluded that the form of the law of emergence given in equation~\eqref{EP_NEQ_EQ14} will hold for any modified gravity theories for which the horizon entropy can be effectively expressed in the form $S=\frac{A}{4G_{\scaleto{\text{eff}}{3.5pt}}}$. This result is similar to the generalized law of emergence (28 in~\cite{https://doi.org/10.48550/arxiv.2111.00726}) derived from non-equilibrium thermodynamics in $f(R)$ gravity, using a generic method, that holds for any modified gravity theories with the above entropy form.

We have analyzed the consistency of the law of emergence~\eqref{EP_NEQ_EQ14}, with the generalized second law and the entropy maximization principle in the context of the non-equilibrium perspective with Tsallis entropy. The validity of generalized second law $\dot{S}\geq0$ implies $N_{\text{sur}}-N_{\text{bulk}}\geq0$ and a constraint on the Tsallis parameter as $\beta<\frac{n+1}{n-1}$. Along with this constraint, the validity of the entropy maximization principle $\ddot{S}<0$ in the final stage of evolution, implies $\dot{N}_{\text{sur}}-\dot{N}_{\text{bulk}}<0$, which demands $((n-1)\beta -1)\dot{\tilde{r}}^{2}_{\scaleto{A}{3.5pt}}<-\tilde{r}_{\scaleto{A}{3.0pt}}\ddot{\tilde{r}}_{\scaleto{A}{3.5pt}}$. It is also clear that these general constraints on the law of emergence in the case of Tsallis modified gravity are similar to those in Einstein, Gauss-Bonnet, and Lovelock gravity theories~\cite{https://doi.org/10.48550/arxiv.2002.02121}. 
\acknowledgments
M Dheepika is thankful to CSIR for the financial support through JRF. Hassan Basari V. T. is thankful to Cochin University of Science and Technology for the financial support through SRF.

\appendix

\section{Gravitational field equation and Friedmann equation based on Tsallis entropy from equilibrium thermodynamic perspective}
\label{a1}
In this section, we would like to derive the field equation and obtain the Friedmann equation from it, based on Tsallis entropy from the equilibrium Clausius relation using the approach of Jacobson~\cite{PhysRevLett.75.1260}, Eling et al.~\cite{PhysRevLett.96.121301}, Asghari and Sheykhi~\cite{10.1093/mnras/stab2671}, and Gennaro~\cite{DiGennaro2022}. Here we consider $(3+1)-$dimensional universe for convenience. Following this approach, the heat flow across the horizon $\mathcal{H}$ is expressed as~\cite{PhysRevLett.75.1260}
\begin{equation}\label{aeq:1}
\delta Q= -\kappa \int_{\mathcal{H}} \lambda T_{\mu \nu}k^{\mu}k^{\nu}d\lambda dA,
\end{equation}
where $\kappa$ is the surface gravity, $T_{\mu \nu}$ is the matter energy-momentum tensor, $k^{\mu}$ is the tangent vector to the horizon generators for an affine parameter $\lambda$, and $dA$ is the area element on a cross-section of the horizon. Assuming the Tsallis entropy as the horizon entropy, $dS_{\scaleto{A}{3.5pt}}$ can be written as~\cite{PhysRevLett.75.1260,10.1093/mnras/stab2671}
\begin{equation}\label{aeq:2}
dS_{\scaleto{A}{3.5pt}}
=\gamma \beta \int_{\mathcal{H}}-\lambda R_{\mu \nu}A^{\beta-1}k^{\mu}k^{\nu}d\lambda dA,
\end{equation}
where $R_{\mu \nu}$ is the Ricci curvature tensor.
Substituting the equations~\eqref{aeq:1} and~\eqref{aeq:2} into the equilibrium Clausius relation $\delta Q=\tilde{T}_{\scaleto{A}{3.5pt}}dS_{\scaleto{A}{3.5pt}}$, we get~\cite{10.1093/mnras/stab2671}
\begin{equation}\label{aeq:3}
-\kappa \int_{\mathcal{H}} \lambda T_{\mu \nu}k^{\mu}k^{\nu}d\lambda dA=\frac{\kappa}{2\pi}\gamma \beta \int_{\mathcal{H}}-\lambda R_{\mu \nu}k^{\mu}k^{\nu}A^{\beta-1}d\lambda dA.
\end{equation}
This can be also written as
\begin{equation}\label{aeq:4}
\int_{\mathcal{H}}\left(- \lambda T_{\mu \nu} + \lambda \frac{\gamma \beta }{2\pi } R_{\mu \nu}A^{\beta-1}\right)k^{\mu}k^{\nu}d\lambda dA=0
\end{equation}
For all null $k^{\mu}$, we have
\begin{equation}\label{aeq:5}
- T_{\mu \nu} + \frac{\gamma \beta }{2\pi }  R_{\mu \nu}A^{\beta-1}=fg_{\mu \nu},
\end{equation}
where $f$ is a scalar. Then imposing the energy-momentum conservation $\nabla^{\mu} T_{\mu \nu}=0$ results in
\begin{equation}\label{aeq:6}
\nabla^{\mu}( \frac{\gamma \beta }{2\pi }  R_{\mu \nu}A^{\beta-1}-fg_{\mu \nu})=0.
\end{equation}
Taking the covariant derivative and using Bianchi identity, we obtain
\begin{equation}\label{aeq:7}
\frac{\gamma \beta }{4\pi }  (\partial_{\nu}R)A^{\beta-1}+\frac{\gamma \beta }{2\pi }R_{\mu \nu}	\partial^{\mu}A^{\beta-1}=\partial_{\nu}f.
\end{equation}
The left-hand side of the equation is not a gradient of the scalar, revealing that the Clausius relation does not hold due to non-equilibrium thermodynamics. However, if we assume the horizon area to be constant~\cite{DiGennaro2022}, that is, $\partial^{\mu}A^{\beta-1}=0$, then we have
\begin{equation}\label{aeq:8}
f=\frac{\gamma \beta }{4\pi }RA^{\beta-1}.
\end{equation}
Substituting the expression for $f$ in equation~\eqref{aeq:5} will result in
\begin{equation}\label{aeq:9}
- T_{\mu \nu} + \frac{\gamma \beta }{2\pi }  R_{\mu \nu}A^{\beta-1}=\frac{\gamma \beta }{4\pi }RA^{\beta-1}g_{\mu \nu}.
\end{equation}
Rearranging the above equation will result in
\begin{equation}\label{aeq:10}
\left(R_{\mu \nu}-\frac{1}{2\pi }Rg_{\mu \nu}\right)A^{\beta-1}=\frac{2 \pi}{\gamma \beta}T_{\mu \nu},
\end{equation}
which is the field equation obtained from the equilibrium Clausius relation when the horizon entropy is assumed to be Tsallis entropy. The corresponding Friedmann equation can be derived in the background of a flat FLRW  $(3+1)$ universe from the time component of the field equation as follows.
\begin{equation}\label{aeq:11}
\left(R_{00}-\frac{1}{2\pi }Rg_{00}\right)(4 \pi \tilde{r}^{2}_{\scaleto{A}{3.5pt}} )^{\beta-1}=\frac{2 \pi}{\gamma \beta}T_{00}.
\end{equation}
On further substitution ($R_{00}=\frac{-3\ddot{a}}{a}$, $R=\frac{6(k+a\ddot{a}+\dot{a}^2)}{a^2}$, $g_{00}=-1$, and $T_{00}=\rho$), we obtain
\begin{equation}\label{aeq:12}
3\left(H^2+\frac{k}{a^2}\right)^{2-\beta}=\frac{2 \pi}{\gamma \beta (4 \pi)^{\beta-1}}\rho.
\end{equation}
Assuming $\gamma=\frac{ (4 \pi)^{1-\beta}}{4\beta G}$, we obtain 
\begin{equation}\label{aeq:13}
\left(H^2+\frac{k}{a^2}\right)^{2-\beta}=\frac{8 \pi G}{3}\rho.
\end{equation}
This is identical to the Friedmann equation obtained by Sheykhi~\cite{SHEYKHI2018118} from the unified first law of thermodynamics using the equilibrium approach by assuming $\gamma=\frac{ (2-\beta)(4 \pi)^{1-\beta}}{4\beta G}$. The same Friedmann equation can be derived from the Clausius relation using Sheykhi's equilibrium approach by assuming $\gamma=\frac{ (2-\beta)(4 \pi)^{1-\beta}}{4\beta G}$. However, here we have derived the Friedmann equation~\eqref{aeq:13} by assuming $\gamma=\frac{ (4 \pi)^{1-\beta}}{4\beta G}$ from the field equation that is obtained from the Clausius relation using Jacobson's approach~\cite{PhysRevLett.75.1260}. The form of the $\gamma$ is different in both equilibrium approaches. The non-equilibrium approach used in section~\ref{s3} is true for any form of the $\gamma$.
\section{Tsallis modified gravity with curvature corrections from non-equilibrium thermodynamic perspective}
\label{TMG_FE}
Considering Tsallis entropy as the entropy associated with the horizon and using the non-equilibrium extension of the Clausius relation, Asghari and Sheykhi~\cite{10.1093/mnras/stab2671} have derived the corresponding gravitational field equation as follows
\begin{equation}\label{FE_NEQ2}
R_{\mu \nu}A^{\beta-1}-\nabla_{\mu}\nabla_{\nu}A^{\beta-1}-\frac{1}{2}RA^{\beta -1}g_{\mu\nu} + \Box A^{\beta-1}g_{\mu \nu}=\frac{2\pi}{\gamma \beta}T^{\text{(m)}}_{\mu \nu}.
\end{equation}
Here $\Box=\nabla^{\alpha}\nabla_{\alpha}$ denotes the covariant d'Alembertian operator. Comparing the above equation with Einstein's field equation, it is clear that the curvature part of the field equation is modified. The extra terms other than the Einstein tensor on the left-hand side of the modified field equation~\eqref{FE_NEQ2} can be absorbed into the energy-momentum tensor on the right-hand side of the equation, then the field equation will be of the form 
\begin{equation}\label{MFE_TE_EEMT}
R_{\mu \nu}-\frac{1}{2}Rg_{\mu\nu}=8\pi G_{\scaleto{\text{eff}}{3.5pt}}\left[\frac{T^{\text{(m)}}_{\mu \nu}}{\beta}+ \frac{\gamma \nabla_{\mu}\nabla_{\nu}A^{\beta -1}}{2\pi}-\frac{\gamma \Box A^{\beta -1}g_{\mu\nu}}{2\pi}\right].
\end{equation}
Though it is evident from the equation~\eqref{FE_NEQ2} that the effects of Tsallis entropy are on the curvature  part of the field equation, the equation~\eqref{MFE_TE_EEMT} shows that it is possible to hide the effects of Tsallis entropy into effective energy-momentum tensor and $G_{\scaleto{\text{eff}}{3.5pt}}$ by absorbing the curvature corrections to the matter part of the field equation. We have followed the latter way to express the effects of Tsallis entropy on the universe for the following reason. The first modified Friedmann equation for a flat $(3+1)$-dimensional universe using the field equations in Tsallis modified gravity~\eqref{FE_NEQ2} is given by
\begin{equation}\label{MFE_TMG}
H^{4-2\beta}=\frac{1}{4\beta-3}\frac{8\pi G}{3}\rho.
\end{equation}
Here $\rho$ is the energy density of the matter and energy content in the universe and $\gamma=\frac{(4\pi)^{1-\beta}}{4G\beta}$. We can obtain the law of emergence in Tsallis modified gravity from the first modified Friedmann equation~\eqref{MFE_TMG}. The above equation can be written as
\begin{equation}\label{MFE_LOE:EQ1}
\tilde{r}_{\scaleto{A}{3.0pt}}^{2\beta-4}=\frac{1}{4\beta-3}\frac{8\pi G}{3}\rho,
\end{equation}
Multiply both sides of equation~\eqref{MFE_LOE:EQ1} by $a^2$ and then differentiate with respect to time. The resulting equation is then divided throughout by $2a\dot{a}$, and it leads to
\begin{equation}\label{MFE_LOE:EQ2}
\left(2\beta-4\right)\tilde{r}^{2\beta-5}_{\scaleto{A}{3.5pt}}\frac{\dot{\tilde{r}}_{\scaleto{A}{3.5pt}}}{2H}+\tilde{r}^{2\beta-4}_{\scaleto{A}{3.5pt}}=\frac{8\pi G}{3(4\beta-3)}\left(\frac{\dot{\rho}}{2H}+\rho\right).
\end{equation}
This equation can be further simplified by multiplying both sides with $8 \gamma \beta (4 \pi \tilde{r}^{2}_{\scaleto{A}{3.5pt}})^{\beta}$ and using continuity equation $\dot{\rho}+3H(\rho+P)=0$, we get
\begin{equation}\label{MFE_LOE:EQ3}
4\gamma  \frac{d\tilde{V}}{dt}=\tilde{r}_{\scaleto{A}{3.0pt}}H \left[\frac{8 \gamma \beta \tilde{A}}{4-2\beta}+\frac{32\pi^2 \tilde{r}^{4}_{\scaleto{A}{3.5pt}}}{  3(4-2\beta)(4\beta-3)}\left(\rho+3P\right)\right].
\end{equation}
Here, $\tilde{A}=(4\pi \tilde{r}^{2}_{\scaleto{A}{3.5pt}})^{\beta}$ and $\frac{d\tilde{V}}{dt}=\frac{\tilde{r}_{\scaleto{A}{3.0pt}}}{2}\frac{d\tilde{A}}{dt}$ are the effective area and the rate of change of effective volume in $(3+1)$-dimensional universe. In order to obtain the law of emergence, we will have to identify the first term on the right-hand side of the above equation as the surface degrees of freedom, that is, $N_{\text{sur}}=\frac{8 \gamma \beta \tilde{A}}{4-2\beta}$. If we substitute for $\gamma=\frac{ (4 \pi)^{1-\beta}}{4\beta G}$ as given in reference~\cite{10.1093/mnras/stab2671}, we obtain $N_{\text{sur}}=\frac{2 \tilde{A}}{4-2\beta}$. The second term on the right-hand side of the above equation should be identified as the bulk degrees of freedom, that is, $N_{\text{bulk}}=-\frac{32\pi^2 \tilde{r}^{4}_{\scaleto{A}{3.5pt}}}{  3(4-2\beta)(4\beta-3)}\left(\rho+3P\right)$.
Even if we multiply equation~\eqref{MFE_LOE:EQ2} with $16 \pi \gamma \tilde{r}^{2}_{\scaleto{A}{3.5pt}}$ and proceed further we will obtain
\begin{equation}\label{MFE_LOE:EQ4}
\frac{dV}{dt}=\tilde{r}_{\scaleto{A}{3.0pt}}H G\left[\frac{2 A}{(4-2\beta)G}+\frac{16\pi^2 \tilde{r}^{4}_{\scaleto{A}{3.5pt}}}{  3}\left(\rho+3P\right)\frac{2\tilde{r}^{2-2\beta}_{\scaleto{A}{3.5pt}}}{(4-2\beta)(4\beta-3)}\right].
\end{equation}
To obtain the law of emergence, we must identify the first term on the right-hand side of the above equation as the surface degrees of freedom, that is, $N_{\text{sur}}=\frac{2 A}{(4-2\beta)G}$. The second term on the right-hand side of the above equation should be identified as the bulk degrees of freedom, that is, $N_{\text{bulk}}=-\frac{16\pi^2 \tilde{r}^{4}_{\scaleto{A}{3.5pt}}}{  3}\left(\rho+3P\right)\frac{2\tilde{r}^{2-2\beta}_{\scaleto{A}{3.5pt}}}{(4-2\beta)(4\beta-3)}$. Thus, keeping the curvature corrections due to Tsallis entropy intact and proceeding further to obtain the law of emergence results in newly defined bulk degrees of freedom, unlike in the existing literature.

\end{document}